\documentclass[%
preprint,   % comment this out for arxiv
preprintnumbers,   % comment this out for arxiv
 amsmath,amssymb,
]{revtex4-2}

\usepackage{graphicx}% Include figure files
\usepackage{dcolumn}% Align table columns on decimal point
\usepackage{bm}% bold math
\newcommand{\Ymax}{Y_{\rm max}}
\newcommand{\lnnavg}{\ln\langle n \rangle}
\newcommand{\navg}{\langle n \rangle}

\begin{document}

%\preprint{APS/123-QED}

\title{Entropy and multiplicity of hadrons in the high energy limit within dipole cascade models}% Force line breaks with \\
%\thanks{}%

\author{Krzysztof Kutak}
\affiliation{Institute of Nuclear Physics Polish Academy of Sciences,\\ Radzikowskiego 152, 31-342 Kraków, Poland}
\affiliation{CPHT, CNRS, Ecole Polytechnique, Institut Polytechnique de Paris, 91120 Palaiseau, France}

\author{S\'andor L\"ok\"os}
\affiliation{Institute of Nuclear Physics Polish Academy of Sciences, Radzikowskiego 152, 31-342 Kraków, Poland}
\affiliation{Hungarian University of Agriculture and Life Sciences, Institute of Technology, Gyöngyös H-3200, Hungary}

\date{\today}

\begin{abstract}
We investigate and compare QCD dipole cascade models, the 1D Mueller dipole model, its high energy limit and its generalization that follows from studies of 1D systems with conformal symmetry. To address the ambiguity stemming from different definitions of the rapidity ranges in experimental measurements, we propose the entropy as the function of the logarithm of the average multiplicity, $S(\lnnavg)$, as a universal observable. From the solutions of the models, we calculate both the entropy and the average charged particle multiplicity and compare to data measured in proton-proton collisions. We obtained these quantities directly from the measured multiplicity distributions and determine the model parameters via fits. We find that the generalized dipole model provides a significantly better description of the data than the 1D Mueller model.

\end{abstract}

\maketitle

%%%%%%%%%%%%%%%%%%%%%%%%%%%%%%%%%%%%%%%%%%%%%%%%%%%%%%%%%%%%%%%%%%%%
\section{Introduction}
\label{sec:intro}

Recent theoretical proposals draw renewed attention to charged particle multiplicity distributions and their Shannon or informational entropy that has a conjectured relation to the entanglement or von Neumann entropy of the initial state partonic system \cite{Kharzeev:2017qzs,Tu:2019ouv,Kutak:2023cwg,Hentschinski:2021aux,Hentschinski:2022rsa,Hentschinski:2023izh,Baker:2017wtt,Bellwied:2018gck,Kou:2023knx,Moriggi:2024tiz,Liu:2022ohy,Liu:2022hto,Liu:2022qqf,Liu:2022bru,Kharzeev:2021nzh,Amorosso:2024glf,Berges:2017hne}. The conjecture also states that in the high energy limit the initial partonic system is maximally entangled. This means that all the partonic microstates have equal probability, and that the system of partons is well characterized with their uniform distribution that maximizes entropy implying a simple formula: $S_{\rm init}=\lnnavg$, where $\navg$ is the average number of the possible partonic microstates. Utilizing patron distribution functions, the average multiplicity of partons, in Deep Inelastic Scattering , can be derived from theory as the function of Bjorken-$x$ and virtuality, $Q^2$, therefore entropy $S(x,Q^2)$ can be obtained and compared to data. The conjecture is supported by analysis of $p+p$ multiplicity data \cite{Kharzeev:2017qzs} and Deep Inelastic scattering data both fully inclusive \cite{H1:2020zpd} that was described in Refs. \cite{Hentschinski:2021aux,Hentschinski:2022rsa,Hentschinski:2024gaa} and diffractive in Ref.~\cite{Hentschinski:2023izh}.  Most recently it has been further confirmed by dedicated Monte Carlo study which clearly demonstrates that bulk of entropy is due to initial state effects contained in PDF \cite{Hentschinski:2025pyq}.

For other related works on entanglement entropy in the context of high energy collisions see Refs.~\cite{Peschanski:2012cw,Kovner:2015hga,Kovner:2018rbf,Dumitru:2023fih,Dumitru:2023qee,Dumitru:2025bib,Dumitru:2022tud,Peschanski:2019yah,Ehlers:2022oal,Ehlers:2022oal,Stoffers:2012mn,Fucilla:2025kit,Lin:2025eci}.

If the hadron--parton duality \cite{RevModPhys.60.373} is true then the final state multiplicities should also be maximal entropy distributions. Recent work showed that the observed distributions in proton-proton collisions are compatible with the idea of the application of the maximal entropy principle \cite{PhysRev.106.620,PhysRev.108.171,Szwed:1987vj} in the final state \cite{Lokos:2025cbu}. It was also observed that the shape of the final state multiplicity distributions is different if they are measured in disjoint, equally wide pseudorapidity intervals (e.g. LHCb~\cite{LHCb:2014wmv} and H1~\cite{H1:2020zpd}) or in symmetric windows, centered to mid-rapidity with increasing width (e.g. ALICE~\cite{ALICE:2015olq,ALICE:2017pcy}, ATLAS~\cite{ATLAS:2010jvh,ATLAS:2016qux,ATLAS:2016zkp}, CMS~\cite{CMS:2010qvf}, UA5~\cite{UA5:1988gup}). However, while the distributions can be derived from the principle of maximum entropy, the comparison with data measured using different rapidity-range definitions requires extra attention.

In this paper our main focus is to compare entropies from different data sets with theoretical models. We show that such comparison is feasible if the entropies are measured as the function of the logarithm of the average number of charged hadrons, i.e., $S(\lnnavg)$ where
\begin{align}
    \navg {=} \sum_n n P(n) \hspace{0.3cm}{\rm and }\hspace{0.3cm} S_h {=}{-}\sum_n P(n) \ln(P(n)).
\label{eq:basic_defs}
\end{align}
It also follows from earlier observations that in the high energy limit, the entropy of the partonic system can be well approximated by the assumption of maximal entropy. The $\lnnavg$ is a beneficial variable because it helps to avoid the ambiguity stemming from the choice of the rapidity variable. Moreover, the proposed $S(\lnnavg)$ entropy function is straightforward to determine from experimentally measured $P(n)$ distributions directly based on Eq.~\eqref{eq:basic_defs}. We obtain the $S(\lnnavg)$ from available $P(n)$s measured in $p+p$ collisions \cite{LHCb:2014wmv, UA5:1988gup, ALICE:2015olq, ALICE:2017pcy, ATLAS:2010jvh,ATLAS:2016qux,ATLAS:2016zkp, CMS:2010qvf} and compare them to theoretical calculations such as the 1D Mueller dipole model \cite{Mueller:1994gb,Levin:2003nc,Kharzeev:2017qzs} and one of its extension \cite{Caputa:2024xkp}. Similar discussion can be found in Refs. \cite{Rybczynski:2025ccy}.

%%%%%%%%%%%%%%%%%%%%%%%%%%%%%%%%%%%%%%%%%%%%%%%%%%%%%%%%%%%%%%%%%%%%
\section{Dipole cascade models}
\label{sec:models}

We utilize two dipole models to address description of the data: the 1D Mueller dipole model \cite{Mueller:1994gb,Levin:2003nc,Kharzeev:2017qzs} in a slightly extended form that was discussed in Refs.~\cite{Hentschinski:2022rsa,Hentschinski:2024gaa}, and one of its generalization \cite{Caputa:2024xkp} that with an extra parameter provides more flexibility.

\subsection{1D Mueller dipole model}

A simple form of a dipole cascade equation that only depends on rapidity, but does not on size (hence it is referred to as 1D Mueller model) can be found in Refs.~\cite{Mueller:1994gb,Levin:2003nc,Kharzeev:2017qzs} and it reads
\begin{align}
    \partial_y P_n(y) = -\alpha n P_n(y) + (n-1)\alpha P_{n-1} (y) ,
\label{eq:Mueller_model}
\end{align}
where the parameter $\alpha$ represents the gluon emission kernel and characterizes the speed of the Balitsky--Fadin--Kuraev--Lipatov (BFKL) cascade evolution \cite{Balitsky:1978ic,Kuraev:1977fs}. One may use  $\alpha=4\bar\alpha_s\ln 2$, in the saddle point approximation. The equation describes the depletion of the probability to find $n$ dipoles due to the splitting into $(n + 1)$ dipoles and the growth due to the splitting of $(n - 1)$ dipoles into $n$ dipoles, it's generalization to account for dipoles merging and transition to vacuum can be found in \cite{Kutak:2025syp}. A solution of this equation \cite{Kharzeev:2017qzs,Mueller:1994gb} can be written as
\begin{align}
    P_n(y)=\frac{1}{C}e^{-\alpha y}\left(1-\frac{1}{C}e^{-\alpha y}\right)^{n-1},
\label{eq:MartinSol}
\end{align}
where the parameter $C$, that slightly generalized the 1D Mueller solution, was introduced in Ref.~\cite{Hentschinski:2022rsa}. We note that the distribution given in Eq.~\eqref{eq:MartinSol} is the geometric distribution with $p=\frac{1}{C}e^{-\alpha y}$ with its support starting at $n=1$,
\begin{align}
    P^{\rm geom}_n(y)=p\left(1-p\right)^{n-1}.
\label{eq:geom_dist}
\end{align}
The $C$ parameter can be interpreted as normalization but we keep it as a free parameter.
The mean multiplicity of dipoles, hence, can be expressed as the function of rapidity
\begin{align}
    \navg = \frac{1}{p} = Ce^{\alpha y},
\end{align}
therefore the rapidity depends on the logarithm of the mean multiplicity,
\begin{align}
    y=\frac{1}{\alpha}\ln\left( \frac{\navg}{C} \right).
\label{eq:rap_navg}
\end{align}
By applying Eq. (\ref{eq:basic_defs}), a relation between the mean multiplicity and the entropy \cite{Kharzeev:2017qzs} can be obtained as:
\begin{align}
    S(y) = \navg \ln \navg - (\navg - 1)\ln(\navg - 1).
\label{eq:Sgeom}
\end{align}
In the sub-asymptotic limit with $C=1$ the  Eq.~\eqref{eq:rap_navg} can be approximated by a so called universal formula \cite{Hentschinski:2023izh}
\begin{align}
    S(\navg) \approx \ln(\navg)+1.
\label{eq:universal}
\end{align} 

\subsection{Generalized Mueller dipole model}
A new cascade equation is 
\begin{align}
    \partial_y P_n(y){=-}\alpha (n{+}2h) P_n(y) {+}\alpha (n{-}1{+}2h)P_{n-1} (y)
\label{eq:CK}
\end{align}
which is a generalization of Eq.~\eqref{eq:Mueller_model} with an extra parameter $h$. The equation has been obtained in the context of study of Krylov complexity \cite{Nandy:2024evd,Rabinovici:2025otw,Baiguera:2025dkc} for the coherent states of the conformal SL(2,R) group and mapped to QCD in Ref.~\cite{Caputa:2024xkp}. Here $h$, is a conformal weight of this group which tels how the operators of the theory scale. This mapping is possible since the BFKL has been shown to obey conformal symmetry and in particular the conformal symmetry has been used to provide its solution in momentum  \cite{Balitsky:1978ic} and impact parameter space \cite{Lipatov:1985uk}. In fact the leading eigenfunction of the BFKL kernel has conformal weight $h=1/2$.
In our setup, the parameter $h$ measures the deviation from the 1D Mueller dipole model. While the Eq.~\eqref{eq:CK} can be rewritten in such form that after setting $h=1/2$ and rescaling (see appendix~\ref{sec:AppRescale}) it reduces to Eq.~\eqref{eq:Mueller_model} in our studies, we use a version of this model first introduced in Ref.~\cite{Caputa:2024xkp}, since in this formulation one can also account for $p_0$, which effectively corresponds to the vacuum contribution. Such contributions were accounted for in recent studies \cite{Hentschinski:2024gaa,Kutak:2025syp} and allowed for successful description of hadronic entropy data. 

The solution of this generalized equation with the previously introduced normalization is given as %\cite{Caputa:2024xkp}

\begin{align}
    P_n(y)=\frac{\Gamma(2h+n)}{n!\Gamma(2h)} \left( \frac{1}{C} e^{-\alpha y}\right)^{2h} \left( 1-\frac{1}{C} e^{-\alpha y}\right)^{n}.
\label{eq:NBD_solution}
\end{align}
One can notice once again that with $p = \frac{1}{C} e^{-\alpha y}$ the solution can be identified with the negative binomial distribution (NBD), where $2h=k$ is the shape or dispersion parameter of the distribution. An explicit form of the entropy of this distribution is not known (a closed form is given in Ref.~\cite{DBLP:journals/corr/abs-1708-06394}), however, with this simple identification, one can explicitly express the mean multiplicity of dipoles or, in the context of Ref.~\cite{Caputa:2024xkp}, Krylov complexity (that is the measure of the spread of the underlying quantum state in Hilbert space), as the function of rapidity in the following form
\begin{align}
    \langle n\rangle= 2h \left(C e^{\alpha y}-1 \right).
\label{eq:navg_y_NBD}
\end{align}
It is worth to emphasis that NBD is a widely applied parametrization of the measured charged particle multiplicity distributions and hitherto explained by phenomenological arguments, e.g., in Ref.~\cite{Giovannini:1991ve}. Moreover, recently it has been shown that NBD distribution gives very good approximation of the dipole models that account for recombination effects and after introducing suitable normalization for transition to vacuum and saturation \cite{Kutak:2025syp}.

\subsection{Comparisons of the models}

\begin{figure}
    \centering
    \caption{The $P_n$ distributions of the 1D Mueller model and the generalized model for different rapidities. The parameters were set to the fitted values given in Tab.~\ref{tab:CK_data_comparison}.}
    \includegraphics[width=0.7\linewidth]{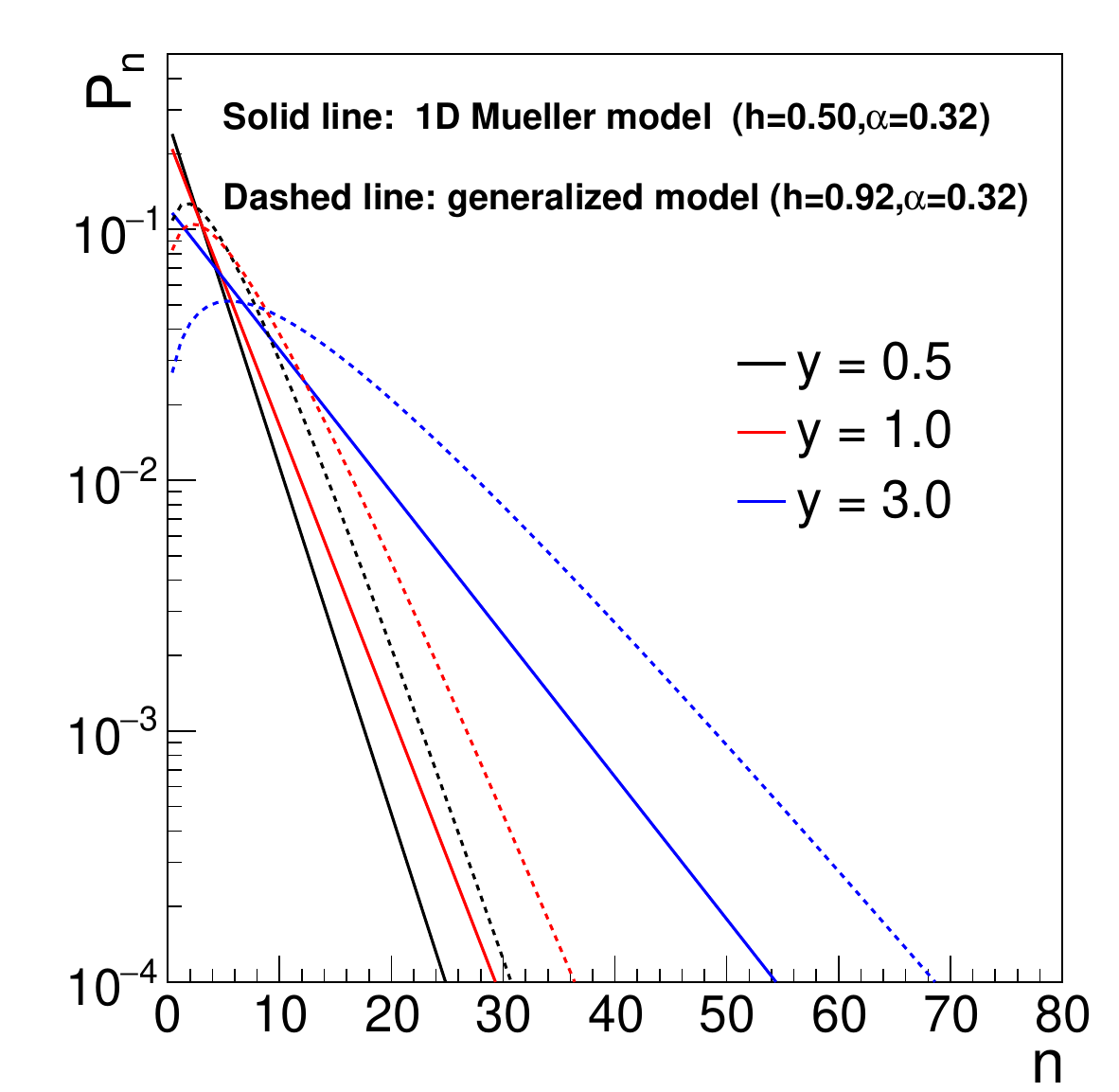}
    \label{fig:rapidity_comparison}
\end{figure}

\begin{figure}
    \centering
    \caption{The log of mean multiplicity as the function of rapidity calculated from the 1D Mueller model and he generalized model. The parameters were set to the fitted values given in Tab.~\ref{tab:CK_data_comparison}.}
    \includegraphics[width=0.7\linewidth]{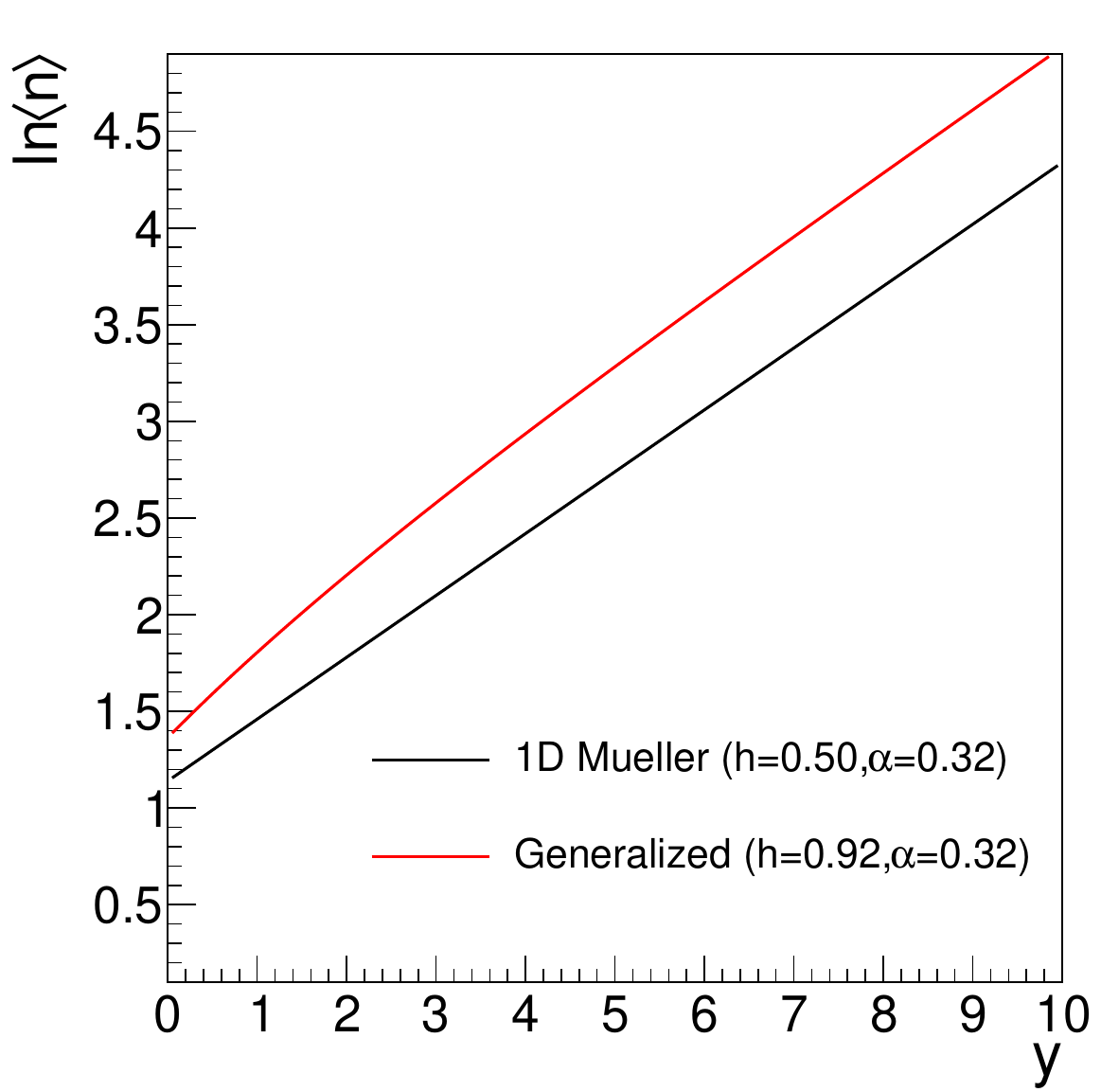}
    \label{fig:y_vs_navg}
\end{figure}

\begin{figure}
\caption{The entropy as the function of rapidity calculated from the 1D Mueller model and he generalized model. The parameters were set to the fitted values given in Tab.~\ref{tab:CK_data_comparison}.}
    \centering
    \includegraphics[width=0.7\linewidth]{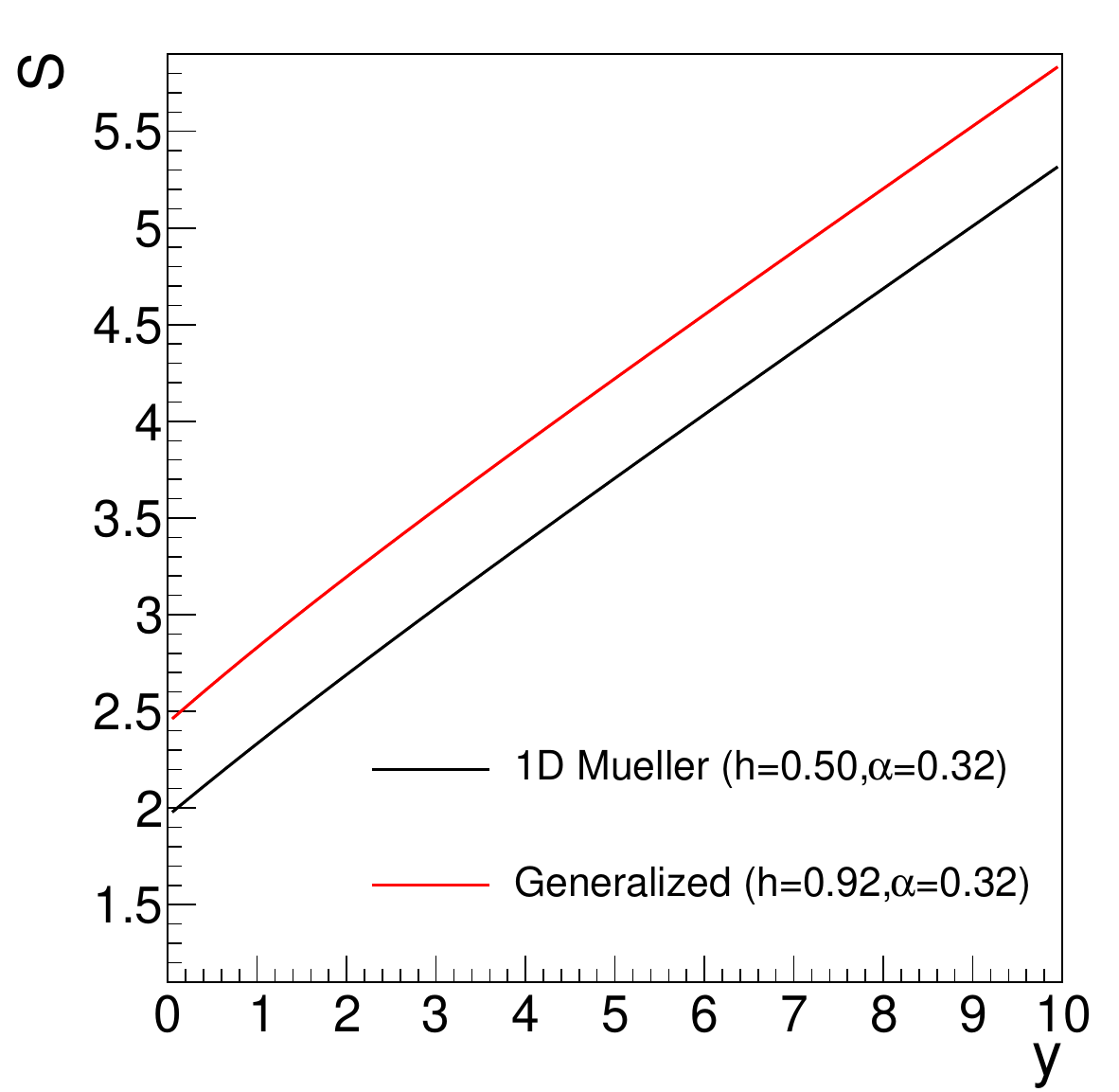}
    \label{fig:S_vs_y}
\end{figure}

We start with the comparison of multiplicity distributions that can be obtained from the two models, as it is shown in Fig.~\ref{fig:rapidity_comparison}. The model parameters for the comparison are taken from our fit results given in Tab.~\ref{tab:CK_data_comparison}. While the multiplicity distributions for the original dipole model features exponential decay, the distributions from the generalized model are more disperse that closely resemble the measured multiplicity data (even in DIS \cite{H1:2020zpd}). Furthermore, they are close in shape to the  multiplicities that can be obtained from the solution of the 3+1D Mueller cascade \cite{Domine:2018myf}. They have the characteristic maximum towards moderate values of $n$. The log of mean multiplicity as the function of rapidity as calculated from the models is shown in Fig.~\ref{fig:y_vs_navg}. Clearly as $h$ increases the mean multiplicity increases as can be also seen from Eq.~\eqref{eq:navg_y_NBD}. The entropy as the function of rapidity, $S(y)$, is shown in Fig.~\ref{fig:S_vs_y}. We see that the generalized model predicts larger entropy for the same value of rapidity. This is a consequence of allowed contribution from vacuum.

%%%%%%%%%%%%%%%%%%%%%%%%%%%%%%%%%%%%%%%%%%%%%%%%%%%%%%%%%%%%%%%%%%%%
\section{Data and fit description}
\label{sec:data_fit_description}

In this section we discuss fits that we obtained utilizing the previously described models. Our main observable is the $S_h(\lnnavg)$ that can be obtained from the models and compared to the data following the original conjecture by \cite{Kharzeev:2017qzs} that $S=S_h$. 
In the following, we will assume that the mean multiplicity of hadrons can be approximated by multiplicity of dipoles and we will use the introduced models to describe hadronic entropy.
We analyzed published charged particle multiplicity distributions $P_n$ at different energies and rapidities, the hadronic entropies $S_h$ were directly obtained as the function of $\lnnavg$. The 1D dipole model  has already been compared to the data \cite{Mueller:1994gb,Levin:2003nc,Kharzeev:2017qzs,Hentschinski:2024gaa,Hentschinski:2022rsa,Hentschinski:2023izh}. Also, the NBD, that is the solution of the generalized model, is a common parametrization of the measured charged particle multiplicity distributions (see e.g. Refs.~\cite{ALICE:2015olq,ALICE:2017pcy,ALICE:2022xip}).

The analyzed data sets are described in Tab.~\ref{tab:data}. However, the psuedorapidity variable is not defined uniquely, by introducing the variable $\lnnavg$, we are able to compare the obtained entropy among experiments. The details of the effect of the rapidity definition are discussed in Ref. \cite{Lokos:2025cbu}. A comprehensive overview of multiplicity distributions and their shape can be found in Ref. \cite{Grosse-Oetringhaus:2009eis}.

We also obtain the uncertainties of $S_h$ and $\lnnavg$. The uncertainty of the entropy was estimated based on Refs.~\cite{PhysRevE.104.024220,doi:10.1137/1104033} and the uncertainty of the $\lnnavg$ was propagated from the uncertainties of the multiplicity distributions. Both uncertainties have to be included into the definition of the $\chi^2$ that is, therefore, given as
\begin{align}
    \chi^2 = \sum_{i=0}^N \frac{(y_i-f(x_i))^2}{ \delta y_i^2 + \left( \frac{\partial f(x_i)}{\partial x}\delta x_i \right)^2} ,
\label{eq:chi2}
\end{align}
where $y_i$ is the obtained $S_h$ at the $i^\textmd{th}$ $\lnnavg$ point, $f(x_i)$ is the model calculation evaluated at the $i^\textmd{th}$ $\lnnavg$, $\delta x_i$ and $\delta y_i$ are the $i^\textmd{th}$ uncertainties of the $\lnnavg$ and $S_h$, respectively. The derivative $\frac{\partial f(x_i)}{\partial x}$ is calculated numerically. The minimization was done by the CERN Minuit2 package.

\begin{table*}
    \caption{The list of data with references that were used in this work.}
    \centering
    \begin{ruledtabular}
    \begin{tabular}{ccccc}
        Experiment & Energies [TeV] & $|\eta|$ width & References & HEPdata entry \\
    \hline
        ALICE & 0.9, 2.76, 7, 8, 13 & 0.5, 0.8, 1, 1.5, 2, 2.4, 3, 3.4, (-3.4 ; 5) & \cite{ALICE:2015olq,ALICE:2017pcy,ALICE:2022xip} & \cite{hepdata.77011,hepdata.78802,hepdata.142463} \\
        ATLAS & 0.9, 2.36, 7, 8, 13 & 2.5 & \cite{ATLAS:2010jvh,ATLAS:2016qux,ATLAS:2016zkp} & \cite{hepdata.57077,hepdata.73012,hepdata.72491} \\
        CMS & 0.9, 2.36, 7 & 0.5, 1, 1.5, 2, 2.4 & \cite{CMS:2010qvf} & \cite{hepdata.57909} \\
        LHCb & 7 & 0.5 (from $\eta=2$ to $\eta=4.5$) & \cite{LHCb:2014wmv} & \cite{hepdata.63498} \\
        UA5 & 0.2, 0.9 & 0.2, 0.25, 0.5, 1, 1.5, 2, 2.5, 3, 3.5, 4, 4.5, 5 & \cite{UA5:1988gup} & \cite{hepdata.15457} \\
    \end{tabular}
    \label{tab:data}
\end{ruledtabular}
\end{table*}

%%%%%%%%%%%%%%%%%%%%%%%%%%%%%%%%%%%%%%%%%%%%%%%%%%%%%%%%%%%%%%%%%%%%
\section{Data comparison}
\label{sec:data_comp}

We compare the $S_h(\lnnavg)$ calculated from the dipole models described in Sec.~\ref{sec:models} and compared them to data detailed in Tab.~\ref{tab:data}. 

\begin{figure}
    \caption{The 1D Mueller model underestimates the entropy in the low average multiplicity region, asymptotically approaches the universal limit discussed around Eq.~\eqref{eq:universal}. The generalized model have similar asymptotics, but is able to describe the data in the low $\lnnavg$ region. The fit parameters are given in Tab. \ref{tab:CK_data_comparison}.}
    \label{fig:data_comparison}
    \centering
    \includegraphics[width=0.7\linewidth]{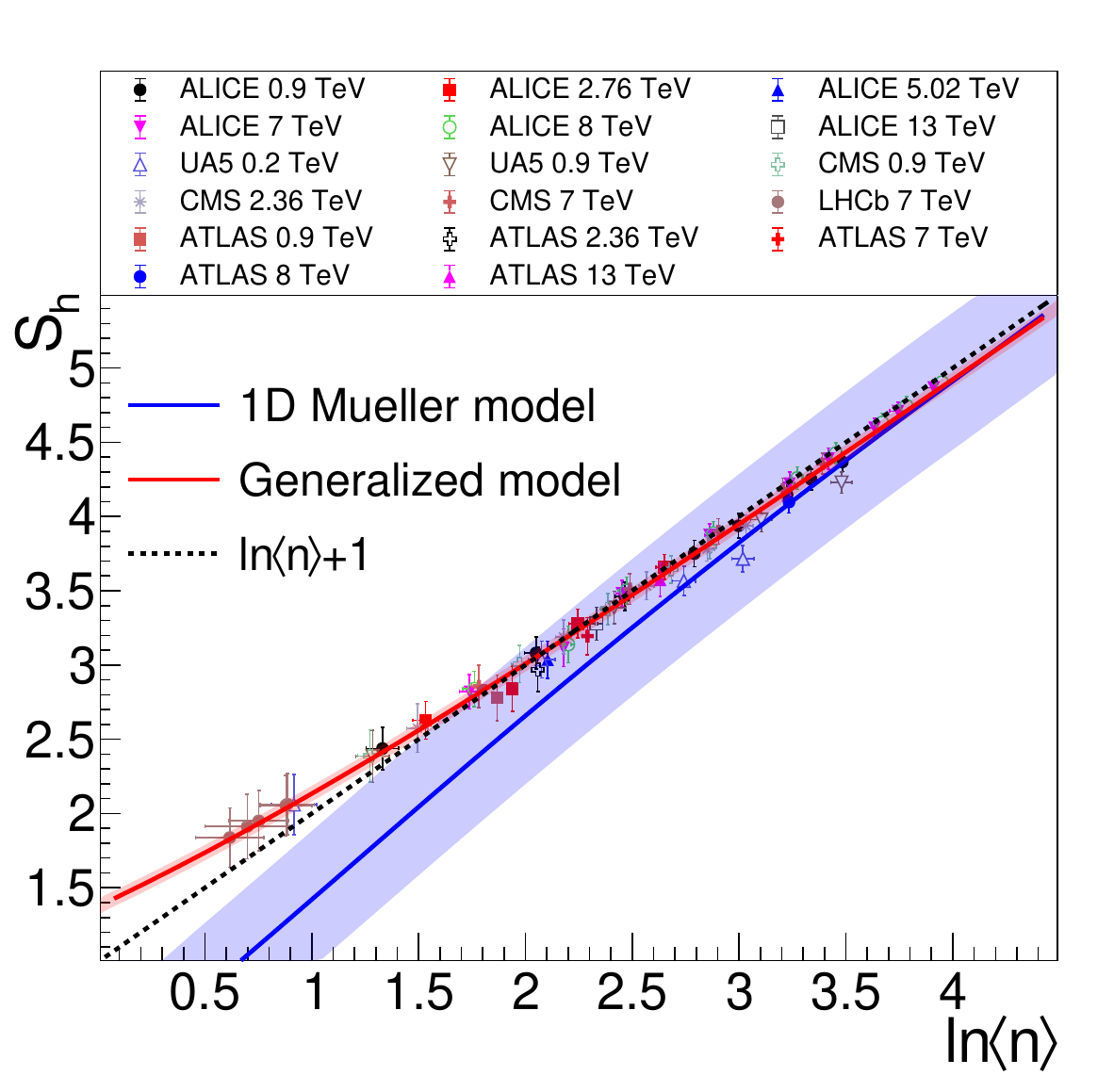}
\end{figure}

The $S(\lnnavg)$ data points were obtained  from measurements at $\sqrt{s_{\rm NN}}=200$ GeV to 13.6 TeV energies and from mid-rapidity to $|\eta|<5$ or $4.5 < \eta < 5$ rapidity ranges. As it was discussed, to get $S(\lnnavg)$, we utilize the definitions in Eq.~\eqref{eq:basic_defs}. However, the obtained entropy is sensitive to the support of the measured multiplicity distributions, that could introduce fluctuations in $S(\lnnavg)$ and could make the comparison between different datasets inconclusive. Hence, first, we perform separate fits to different datasets where it was possible (i.e., at least three $S(\lnnavg)$ points were available). We observed that the obtained fit parameters for each models with different datasets are consistent with each other within statistical uncertainties, despite of the differences in the support of the distributions, therefore we perform a combined fit to all available data that can be seen in Fig.~\ref{fig:data_comparison}.

The 1D dipole model has two, while the generalized model has three parameters. The 1D model was compared to data in Ref.~\cite{Hentschinski:2024gaa}, so we checked that if our fits give compatible values to those are published in Tab. III. of Ref.~\cite{Hentschinski:2024gaa}. We have found a good agreement, therefore we fixed the $\alpha$ parameter to the virtuality-averaged published value.

Based on the fits we conclude that the generalized model gives a statistically acceptable description of the data while the 1D dipole model underestimates it in the low $\lnnavg$ region. It can be also observed that both models go towards the universal limit given in Eq.~\eqref{eq:universal}. Such universal asymptotic behavior was already discussed in Ref.~\cite{Kharzeev:2017qzs}.

The parameter values of the combined fit with their statistical uncertainties and their $\chi^2/$NDF are provided in Tab.~\ref{tab:CK_data_comparison}. It can be seen that the $C$ parameter cannot be constrained well by the fits that results in a considerable uncertainty of the the 1D Mueller model that is exhibits a wide uncertainty band in Fig.~\ref{fig:data_comparison}. The new parameter of the generalized model, however, can be determined precisely, therefore the model uncertainty is small.

The preferred value obtained, $h=0.92$, for the generalized model clearly indicates that the model effectively accounts for effects beyond BFKL. The value and shape of the multiplicity distributions suggest that DGLAP effects may also be effectively taken into account. The calculation of multiplicity for the two models (see Fig.~\ref{fig:y_vs_navg}) shows that the generalized model predicts larger values, which are favored by the experimental data.

\begin{table}[]
    \centering
    \begin{tabular}{|c|c|c|}
    \hline
       Parameter  & 1D dipole & Generalized model \\
    \hline
       $h$      & 0.50 (fixed)  & $0.92\pm0.05$  \\
       $\alpha$ & 0.32 (fixed)  & 0.32 (fixed)  \\
       $C$      & $3.13\pm0.48$ & 3.13 (fixed) \\
       \hline
       $\chi^2/$NDF & 309/63   & 24/63 \\
    \hline
    \end{tabular}
    \caption{The parameters and fit qualities of the dipole models \cite{Mueller:1994gb,Levin:2003nc,Kharzeev:2017qzs,Caputa:2024xkp}. The fixed values are based on earlier results \cite{Hentschinski:2024gaa}.}
    \label{tab:CK_data_comparison}
\end{table}

%%%%%%%%%%%%%%%%%%%%%%%%%%%%%%%%%%%%%%%%%%%%%%%%%%%%%%%%%%%%%%%%%%%%
\section{Conclusions and discussion}
\label{sec:conclusion}
Final state hadronic entropy is an interesting observable that has a conjectured relation to the entropy of the entangled quantum system of the initial state partons; that is the proton. This conjecture gets more support by recent Monte Carlo studies where it has been shown that the bulk of entropy comes from initial state showed to be directly connected to Initial State Radiation (ISR) \cite{Hentschinski:2025pyq}.

In this paper we applied two  dipole models to calculate entropy as the function of the logarithm of the average multiplicity, $S(\lnnavg)$. We obtained the same quantity directly from the data and showed that it can be determined without any methodological ambiguity. We compared the model calculations to the data and demonstrated that the generalized model can describe the data very well.

We note that there is room for improvement on both the experimental and theoretical sides. While the generalized model has an explicit solution, it does not account for potentially relevant effects of recombination that may lead to saturation. Charged-particle multiplicities should also be measured in more detail, across wider rapidity ranges and energies, and in various systems. Future instruments, such as the Electron-Ion Collider, may be able to push the precision limits of such studies; however, existing experiments could also provide interesting results from exotic reactions (e.g. neutrino DIS experiments like FASER or SND) or from more precise measurements of multiplicities in proton–proton collisions at higher energies and forward regions (ALICE FoCal).

%%%%%%%%%%%%%%%%%%%%%%%%%%%%%%%%%%%%%%%%%%%%%%%%%%%%%%%%%%%%%%%%%%%%
\section*{Acknowledgments}
S.L. is grateful for the support of the Miniatura 7 grant of the National Science Centre of Poland and thanks the hospitality and insightful discussions to M. P{\l}osko{\'n} in Berkeley and G. Contreras in Prague. The authors would like to thank the fruitful discussions to P. Caputa,  M. Hentschinski, H. Jung, M. Prasza{\l}owicz,  M. {\v Sumbera}, Z. Tu, D. Kharzeev and J. Otwinowski.

%\bibliography{citation}% Produces the bibliography via BibTeX.
\section*{References}

%%%%%%%%%%%%%%%%%%%%%%%%%%%%%%%%%%%%%%%%%%%%%%%%%%%%%%%%%%%%%%%%%%%
\section*{Appendix}
\appendix

%\section{The maximal entropy principle}
%\input{appendix_A_maxent}
%\label{sec:maxent}

\section{Rescaling the general solution}
\label{sec:AppRescale}

In this appendix we show how the solution of the generalized model in Eq. (9) reduces to the Mueller model.

Let's set $h=1/2$ and $C=1$ in the solution Eq.~\eqref{eq:NBD_solution}, that is then written as
\begin{align}
    P_n(y)=e^{-\alpha y}(1-e^{-\alpha y})^n.
\end{align}
Let's multiply by $(1-e^{-\alpha y})$, so we obtain
\begin{align}
    P_{n+1}(y)=e^{-\alpha y}(1-e^{-\alpha y})^{n+1}.
\end{align}
Now, let's multiply Eq.~\eqref{eq:CK} by $(1-e^{-\alpha y})^{n+1}$, so we obtain
\begin{align}
    \partial_y P_{n+1}(y)=\alpha n P_n(y)-\alpha(n+1)P_{n+1}(y)    
\end{align}
Substituting $n+1=k$, we obtain the Mueller model
\begin{align}
    \partial_y P_{k}(y)=\alpha (k-1) P_{k-1}(y)-\alpha k P_k(y).
\end{align}

\section{Entropy scaling in proton-proton collisions}
\label{sec:Ymax_scaling}

Universal scaling of charged particle multiplicity distributions, the so called Koba--Neilsen--Olesen (KNO) scaling was observed at different energies and in various colliding systems \cite{Koba:1972ng} as
\begin{align}
    \navg P(n) = \psi\left( z \right)
\label{eq:KNO_scaling}
\end{align}
where $\psi$ is the KNO scaling function and its variable $z=\frac{n}{\navg}$. Since the observation of the scaling there were several proposal for the functional form of $\psi$, e.g., Refs.~\cite{deGroot:1975cvq,Hegyi:1996bt,Hegyi:1996ut,Hegyi:1997im}.
If it is substituted into the general entropy formula in Eq.~\eqref{eq:basic_defs}, the following equation can be derived \cite{Kovner:2022jqn,Liu:2022bru} as the direct consequence of the KNO scaling
\begin{align}
    S_h = \ln \langle n \rangle + S_\psi.
\label{eq:Sdiff}
\end{align}
Similar derivation was performed in Ref. \cite{SIMAK1988159}, where a scaling behavior of $S_h/\Ymax$ was shown with maximum rapidity, $\Ymax$. This scaling can be obtained from Eq. \eqref{eq:Sdiff}. If the experimentally observed relation $\langle n \rangle \propto s^a$ is substituted into Eq.~\eqref{eq:Sdiff} as $\ln \langle n \rangle \propto a\ln \left( \sqrt{s}/m_p \right) = a\Ymax$, where the maximum rapidity was defined as $\Ymax=\ln \left( \sqrt{s}/m_p \right)$, the scaling follows as
\begin{align}
    \frac{S_h - S_\psi}{\Ymax} = a
\label{eq:Sratio}
\end{align}
Since the ratio is derived from the KNO-scaling, it is independent from the actual form of the $P(n)$ or $\psi$ as long as they are probability distribution functions. The ratio is approximately independent from $\sqrt{s_{\rm NN}}$ as $\Ymax$ grows.

\end{document}